# Are Quantitative Features of Lung Nodules Reproducible at Different CT Acquisition and Reconstruction Parameters?

Barbaros S. Erdal, Mutlu Demirer, Chiemezie C. Amadi, Gehan F. M. Ibrahim, Thomas P. O'Donnell, Rainer Grimmer, Andreas Wimmer, Kevin J. Little, Vikash Gupta, Matthew T. Bigelow, Luciano M. Prevedello, Richard D. White

*Abstract*— Consistency and duplicability in Computed Tomography (CT) output is essential to quantitative imaging for lung cancer detection and monitoring. This study of CT-detected lung nodules investigated the reproducibility of volume-, density-, and texture-based features (outcome variables) over routine ranges of radiation-dose, reconstruction kernel, and slice thickness. CT raw data of 23 nodules were reconstructed using 320 acquisition/reconstruction conditions (combinations of 4 doses, 10 kernels, and 8 thicknesses). Scans at 12.5%, 25%, and 50% of protocol dose were simulated; reduced-dose and full-dose data were reconstructed using conventional filtered back-projection and iterative-reconstruction kernels at a range of thicknesses (0.6-5.0 mm). Full-dose/B50f kernel reconstructions underwent expert segmentation for reference Region-Of-Interest (ROI) and nodule volume per thickness; each ROI was applied to 40 corresponding images (combinations of 4 doses and 10 kernels). Typical texture analysis metrics (including 5 histogram features, 13 Gray Level Co-occurrence Matrix, 5 Run Length Matrix, 2 Neighboring Gray-Level Dependence Matrix, and 2 Neighborhood Gray-Tone Difference Matrix) were computed per ROI. Reconstruction conditions resulting in no significant change in volume, density, or texture metrics were identified as "compatible pairs" for a given outcome variable. Our results indicate that as thickness increases, volumetric reproducibility decreases, while reproducibility of histogram- and texture-based features across different acquisition and reconstruction parameters improves. In order to achieve concomitant reproducibility of volumetric and radiomic results across studies, balanced standardization of the imaging acquisition parameters is required.

*Index Terms*— computed tomography, dose, reconstruction kernel, slice thickness, lung nodule, volume, histogram, texture, radiomics.

†This project was partially funded by Edward J. DeBartolo, Jr. Family Fund, and NIH Center for Accelerated Innovations. (Corresponding author: barbaros.erdal@osumc.edu).

B. S. Erdal, M. Demirer, C. C. Amadi, G. F. M. Ibrahim, K. J. Little, V. Gupta, M. Bigelow, L. M. Prevedello, and R. D. White are in the Department of Radiology, The Ohio State University College of Medicine, Columbus, OH 43210,USA(email:barbaros.erdal@osumc.edu;mutlu.demirer@osumc.edu;chiemezie.amadi@osumc.edu;gehan.ibrahim@osumc.edu;kevin.little@osumc.edu;vikash.gupta@osumc.edu;matthew.bigelow@osumc.edu;luciano.prevedello@osumc.edu;richard.white@osumc.edu).

T. P. O'Donnell, R. Grimmer, and A. Wimmer are with the Siemens Healthineers, Malvern, PA 19355, USA and Erlangen, Germany (e-mail: tom.odonnell@siemens-healthineers.com;rainer.grimmer@siemens-healthineers.com;andreas.aw.wimmer@siemens-healthineers.com)

## I. INTRODUCTION

LUNG nodules have traditionally been evaluated with two-dimensional (2-D) linear measurements on chest Computed Tomography (CT) (e.g., Response Evaluation Criteria in Solid Tumors (RECIST) [1]). However, three-dimensional (3-D) volumetric assessments of lung nodules are gaining importance due to: 1. improved representation of disease extent and therapeutic-response; and 2. less user-dependency and higher reproducibility of the results [2]. Concurrently, new diagnostic and treatment paradigms increasingly emphasize the value of quantitative radiomic features of lung nodules and surrounding lung tissue as indicators of tumor type, aggressiveness, and/or responsiveness to treatment [3], [4]. The associated quantitative metrics assume adequacy and uniformity in CT data acquisition and reconstruction despite well-recognized wide inter-scan variability in: 1. examination protocoling for image acquisition; 2. image reconstructions and displays; and 3. CT equipment and technologist capabilities and performances. Although the susceptibility of several lung nodule measurements (volumetric or radiomic) to variations in individual CT acquisition parameters has already been recognized based on preliminary studies of clinical or phantom data [5]–[8], the concomitant effects that routine ranges of radiation-dose, reconstruction kernel, and slice thickness have on nodule volume and texture features have not yet been reported.

The purpose of this research was to: 1. determine the impact of a variety of standard imaging acquisition and reconstruction parameters on lung nodule volumes and radiomic features; 2. identify potential imaging acquisition parameters that allow consistency and reproducibility of volumetric and/or radiomic features of lung nodules.

## II. MATERIALS AND METHODS

### A. Acquisition and Reconstruction Parameter Variations

Raw Digital Imaging and Communications in Medicine (DICOM) data from 23 non-contrast-enhanced chest CT

examinations demonstrating a lung nodule were acquired from either single-source (Definition AS, AS Plus, Edge) or dual-source (Definition Flash) multi-detector CT systems [Somatom series: Siemens Healthineers, Forchheim, Germany (https://www.healthcare.siemens.com/computed-tomography)] with ranges of settings (CTDIvol = 1.64-11.34 mGy, Tube voltage = 100-140 kV, Q.ref.mAs = 50-100 mAs, Eff.mAs = 41-154 mAs, and pitch = 0.6-1.0). Based on this data, chest CT data-sets underwent the following:

1. Simulation at different dose levels (simulated 12.5%, 25%, 50%), as well as 100% of total dose of clinical protocols
2. Reproduction with 10 different kernels based on either Filtered Back Projection (FBP) or Iterative Reconstruction (IR) (FBP: B31f, B40f, B50f, B60f, B70f; SAFIRE (strength 3): I26f, I31f, I40f, I50f, I70f) [ReconCT: Siemens Healthineers, Forchheim, Germany]
3. Recreation at 8 different slice thicknesses (0.6, 0.75, 1.0, 1.5, 2.0, 3.0, 4.0, and 5.0 mm) using an offline reconstruction system [ReconCT: Siemens Healthineers, Forchheim, Germany].

Based on combinations of the aforementioned variations in acquisition or reconstruction parameters, 320 versions of each examination (i.e., 4 doses x 10 kernels x 8 thicknesses) were created.

*B. Segmentation and Volume Measurement*

In order to display the nodules to dedicated thoracic radiologists (CCA and GFMI each with 6-8 years of post-fellowship experience), a custom Graphical User Interface (GUI) was built [MeVisLab version 2.8, Windows 64 bit, VS2013: Bremen, Germany (https://www.mevislab.de/)] (Fig. 1). This GUI integrated the functionalities of a commercial nodule segmentation algorithm [Pulmonary Package: Siemens Healthineers, Forchheim, Germany (https://www.healthcare.siemens.com/computed-tomography/options-upgrades/clinical-applications/syngo-inspace-lung-parenchyma)] as they would appear within an established commercial post-processing platform [syngo.via: Siemens Healthineers, Forchheim, Germany (https://www.healthcare.siemens.com/medical-imaging-it/advanced-visualization-solutions/syngovia)].

For each chest CT examination, sets of 100%-dose/FBP B50f-kernel images were reconstructed at the aforementioned 8 different slice thicknesses, with nodule extent and shape previously independently confirmed by consensus between the two dedicated thoracic radiologists. This resulted in 8 reference Region-Of-Interest (ROI) stacks, each corresponding to a specific slice thickness, which were then applied to the 40 different dose-kernel combinations (i.e., 4 doses x 10 kernels) at constant slice thickness (Fig. 2).

*C. Analysis of Reproducibility of Volumetric Measurements*

Using automatic 3-D segmentation, 8 volumetric measurements were obtained for each nodule (Fig. 3.a) corresponding to each slice thickness (Fig. 3.b) and

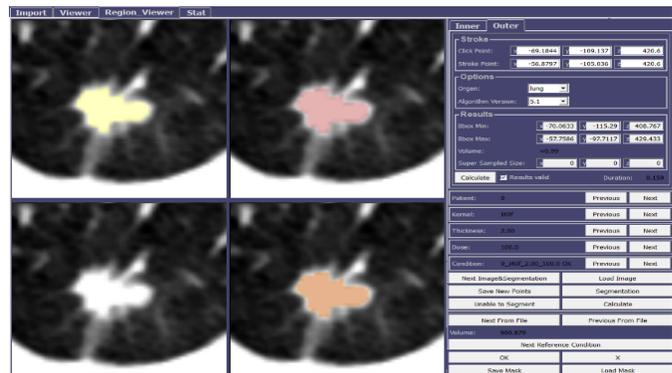

Fig. 1. A custom GUI allowed thoracic radiologists to evaluate nodules. For the purposes of this study, only the solid components were considered. Semi-automated segmentations made by radiologists for a given nodule are shown. Top left in Yellow: Radiologist 1; top right in Red: Radiologist 2; bottom left: non-segmented original; bottom right in Orange, both radiologist's segmentations combined.

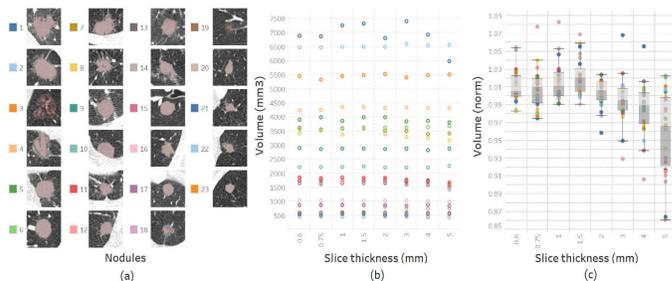

Fig. 2. Segmentation procedure for lung nodules. Each nodule was segmented from 100%-dose/FBP B50f-kernel images reconstructed at 8 different slice thicknesses. The resulting 8 thickness-specific ROI stacks were then applied to the corresponding images reconstructed at 4 different dose levels and 10 different kernels at stable slice thickness.

Fig. 3. The methodologies used during volumetric assessment of lung nodules are shown. The following represent expert segmentation of lung nodules: (a) Reference radiologist-segmented ROI stacks obtained using 100%-dose level, a FBP B50f kernel, and 1-mm slice thickness. Nodules are sorted by their average volumes and numbered 1 to 23. (b) Calculated volumes for each nodule using 8 reference ROIs depending on slice thicknesses (colors indicate the corresponding nodules shown on the left). (c) Normalized volumes of nodules obtained using their average volumes.

normalized by their averages using (1) as follows ($V_i$: nodule volume for i-th thickness):

$$V(norm) = \frac{V_i(mm^3)}{\frac{1}{8}\sum_{k=1}^{8} V_k(mm^3)} \quad (1)$$

Using the distributions of normalized volumes for each slice thickness (Fig. 3.c), t statistics were calculated using 2-tailed t-test (2) to evaluate the compatibilities of slice thicknesses based on volumetric measurements. If $t<1.96$ ($P<0.05$), slice thicknesses are accepted as compatible with 95% confidence interval; $m_1$, $m_2$: means; $s_1$, $s_2$: standard deviations; $n_1$, $n_2$: number of samples (nodules):

$$t = \frac{abs(m_1 - m_2)}{\sqrt{\frac{(s_1)^2}{n_1} + \frac{(s_2)^2}{n_2}}} \quad (2)$$

### D. Image-Feature Extraction

A range of extracted radiomic image features was assessed (Table 1). They included 28 image-texture features (available in MeVisLab) as follows: 1. 5 Histogram-based features; 2. 13 Gray Level Co-occurrence Matrix (GLCM)-based features; 3. 5 Run Length Matrix (RLM)-based features; 4. 2 Neighboring Gray-Level Dependence Matrix (NGLDM)-based features, and 5. 3 Neighborhood Gray-Tone Difference Matrix (NGTDM) based features which were computed for all 320 segmented image volumes of each nodule.

### E. Compatibility of Image Reconstruction Conditions Based on Radiomic Features

Means and standard deviations of the aforementioned 28 image features were computed for each of the 320 segmented images of each nodule. Statistically significant changes in image features were evaluated using 2-tailed t-test. Reconstruction condition pairs that satisfy $t$ values of lower than <1.96 (P<0.05) are accepted as compatible with 95% confidence interval. Compatibility ratios were calculated using Equation 3; $CR_{(R_i,R_j)}$: compatibility ratio of reconstruction condition pair $(R_i, R_j)$; $R_i$: $i$-th reconstruction condition ($i$=1,2,…,320); $C_{(R_i,R_j)}(f,p)$: compatibility of $(R_i, R_j)$ pair for feature $f$ (1,…,28) and patient $p$ (1,…,23) calculated using t-test (Equation 2):

$$CR_{(R_i,R_j)} = \frac{\sum_{f=1}^{28} \sum_{p=1}^{23} C_{(R_i,R_j)}(f,p)}{28 \times 23} \times 100 \quad (3)$$

We produced a compatibility map of reconstruction conditions (Fig. 4) to highlight effects of changes in slice thickness (T), kernel sharpness (K), and dose (D). Reconstruction condition parameters are sorted based on total number of compatibilities. Thickness order: 5mm, 4mm, 3mm, 2mm, 1.5mm, 1mm, 0.75mm, 0.6mm. Kernel order: I26f, I31f, B31f, I40f, B40f, I50f, B50f, I70f, B60f, B70f. Dose level order: 100%, 50%, 25%, 12.5%.

## III. RESULTS

### A. Reproducibility of Volumetric Measurements

Volume comparisons were performed on 8 slice thicknesses on all 23 cases (Fig. 5). The slice thickness resulting in the least amount of volumetric variability was 2-mm, with +0.39%±1.59 (mean±SD) variation from average volume. On the other hand, 1.5-mm thickness gave the highest volumes with +2.16%±2.07, and 5-mm the lowest volumes with -4.22%±4.65. While standard deviations were relatively stable for slice thicknesses below 2mm, increasing thickness beyond 2mm was associated with rapidly increasing standard deviation. Assessment of volumetric reproducibility shows that increasing slice thickness decreases compatibility (Fig. 6).

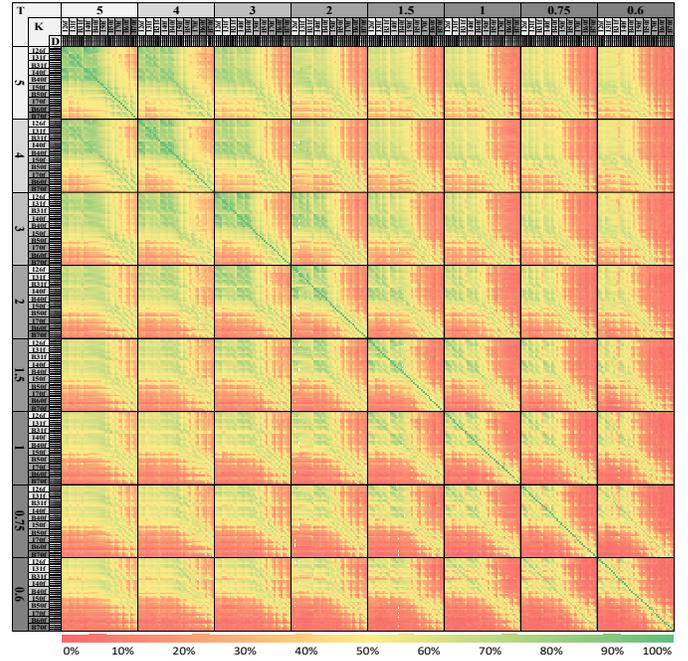

Fig. 4. Reconstruction-condition compatibility map based on extracted features and patients. Intersections of conditions are highlighted (Red: incompatible, Green: compatible) based on their compatibility ratios calculated using t-test. Diagonal shows 100% compatibility which satisfies all comparisons (28x23). Changing reconstruction parameters (thickness/dose/kernel-sharpness) decreases the compatibility. In order to obtain higher compatibility, changes to the reconstruction parameters should be applied carefully. For example if thickness needs to be switched from 2 mm to 0.75 mm, softer kernels and/or higher dose levels are needed as seen from the intersection of two thicknesses.

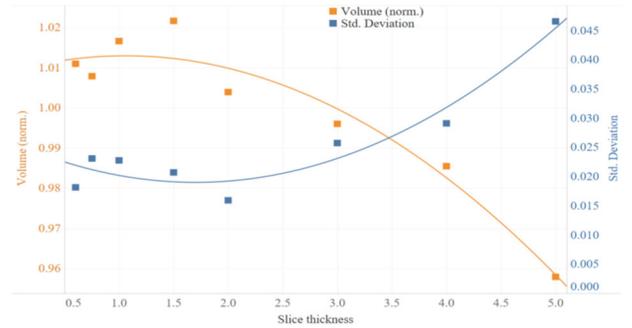

Fig. 5. Normalized volumetric measurements and trend lines based on slice thickness.

| Thickness (mm) | 5.00 | 4.00 | 3.00 | 2.00 | 1.50 | 1.00 | 0.75 | 0.60 |
|---|---|---|---|---|---|---|---|---|
| 5.00 | 1.000 | 0.023 | 0.002 | 0.000 | 0.000 | 0.000 | 0.000 | 0.000 |
| 4.00 | 0.023 | 1.000 | 0.208 | 0.012 | 0.000 | 0.000 | 0.007 | 0.001 |
| 3.00 | 0.002 | 0.208 | 1.000 | 0.227 | 0.001 | 0.007 | 0.118 | 0.031 |
| 2.00 | 0.000 | 0.012 | 0.227 | 1.000 | 0.003 | 0.037 | 0.522 | 0.177 |
| 1.50 | 0.000 | 0.000 | 0.001 | 0.003 | 1.000 | 0.449 | 0.042 | 0.076 |
| 1.00 | 0.000 | 0.000 | 0.007 | 0.037 | 0.449 | 1.000 | 0.208 | 0.367 |
| 0.75 | 0.000 | 0.007 | 0.118 | 0.522 | 0.042 | 0.208 | 1.000 | 0.613 |
| 0.60 | 0.000 | 0.001 | 0.031 | 0.177 | 0.076 | 0.367 | 0.613 | 1.000 |

Fig. 6. P values for compatibility analysis of slice thicknesses based on volumetric measurements. Higher P value indicates higher compatibility. Intersection of compatible thicknesses with P values higher than 0.05 are highlighted with green; P values lower than 0.05 (highlighted with orange to red) indicate incompatible slice thicknesses.

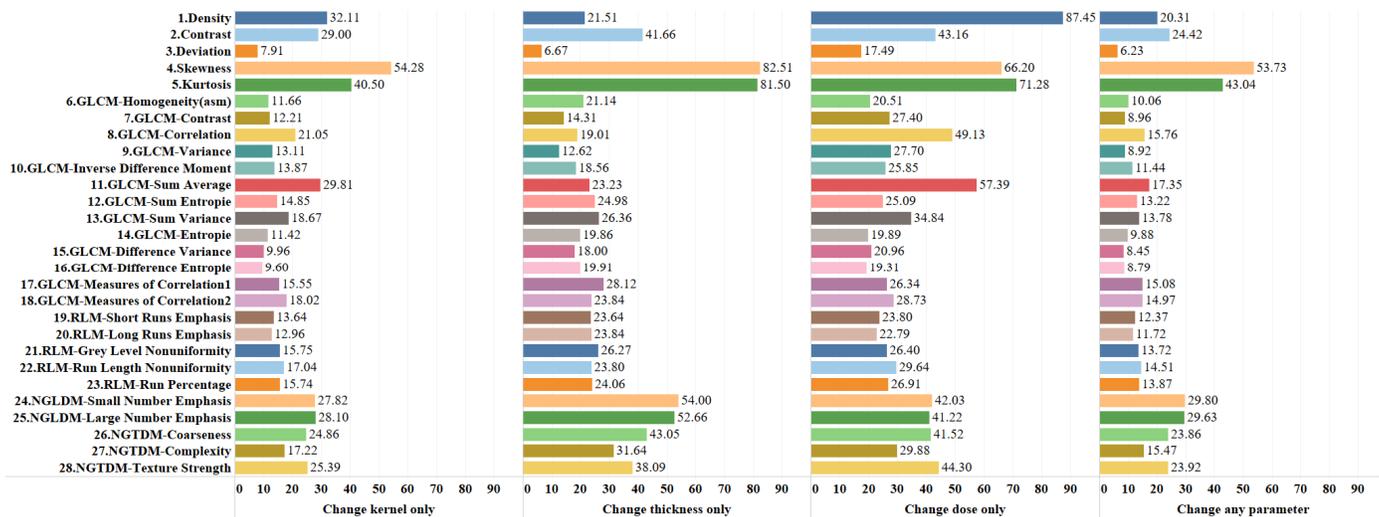

Fig. 7. Reproducibility ratios of features based on dose, kernel, and thickness changes.

B. *Reproducibility of Radiomic Features*

320 reconstruction conditions were compared to each other on 28 texture features and 23 cases, totaling 65,945,600 (320x320x28x23) comparisons. As shown in Fig. 4, highest average compatibility of 24.47% was achieved using the combination of highest slice thickness, smoothest kernel and highest dose level (5mm/I26f/100%). Compatibility decreases while decreasing slice thickness and/or kernel smoothness and/or dose level. Lowest average compatibility of 2.65% was at the combination of lowest slice thickness, sharpest kernel and lowest dose level (0.6mm/B70f/12.5%). Fig. 7 shows percentage compatibilities of radiomic features based on dose, kernel, and thickness changes. Most robust feature was the density against dose changes (87.45% compatibility); and skewness was most robust for kernel and slice thickness condition changes (53.73%-82.51% compatibility). Deviation was the weakest feature for all cases. In average, GLCM based feature group was the most vulnerable feature group (19.41% average compatibility). In addition, results showing percentage compatibilities of conditions based on kernel sharpness, slice thickness and dose levels are presented in Fig. 8, 9 and 10 respectively.

As it can be seen from the figures, while stable volumetric measurements can be obtained (e.g. 2mm slice thickness) and volumetric measurement errors can be predicted when slice thicknesses are changed (Fig. 5); that is not the case for radiomic features (Fig. 6-7). Texture measurements can be very unstable when conditions are altered.

IV. DISCUSSION

Robust image features are vital for designing and standardizing anatomic and radiomic-based diagnosis and prognosis decisions [9]–[13]. In this study, we have investigated the effects of image acquisition and reconstruction conditions on volume and radiomic features derived from chest CT scans of lung nodules. These conditions contained an extensive list of combinations (320 versions of each examination: 4 doses x 10 kernels x 8 thicknesses).

Lung nodule detection is enhanced when thinner CT slices are produced [14]–[16]; screening CT scans are preferentially performed with 1 to 2.5-mm-thick slices [14], [17], as was done in the National Lung Screening Trial [18]. Solid nodules ≥ 4-mm in diameter are currently considered important [14] and were the focus of this research. Accurate and reliable measurement of lung nodule size from CT scans is a key biomarker in the diagnosis of lung cancer; the estimation of nodule growth rates serves as a predictor of malignancy and size change reflects efficacy of a treatment [19], [20].

A related challenge is the consistency in establishing lung nodule size [14]. Reliable sizing of nodules has traditionally been limited by subjectivity in selection of the desired dimensions to measure and non-uniformity in measuring by manual manipulation of digital calipers; this represents a source of disagreement between readers and reference standards [2], [14]. Manually measuring nodule size involves laboriously inspecting all images that include the lesion [19]. To provide a standard method for nodule-size measurement, the RECIST working group proposed the common use of the 1-D maximal diameter as an efficient standard estimator of lesion volume [1], [19], [21]. While such basic mass size measurements are typically used in clinical practice, 3-D volume measurements are growing in importance due to evidence that 3-D volumetry is more robust for quantifying tumor size [20], [22]. Poor agreement between 3-D and simpler methods is commonly seen when the nodule does not conform to spherical or ellipsoidal assumptions that underlie 1-D and 2-D measurements, respectively; in the context of spatial extent for a 3-D object without restrictions on shape, size is best expressed by the volume occupied by the object [19].

Consequently, there is increasing interest in computer-assisted methods aiding the radiologist in measuring the size of lung nodules using volumetric methods [23]–[26], despite the fact that their calibration and validation becomes a new change [19]. Although lung-nodule volumetry has the potential to improve patient management, there is considerable and largely unpredictable variability in its execution related to produced slice thickness, reconstruction algorithm, and scan dose [20], [22], [27], [28]. Our research addressed this issue and demonstrated that while increasing thickness decreases volumetric reproducibility, it improves the reproducibility of histogram- and texture-based features across different acquisition and reconstruction parameters.

Texture analysis is promising for phenotyping and segmenting cancerous tissues [6], [7]. However, Buch et al. [6] highlighted the major looming problem pertaining to radiomics and big data, that despite collecting increasing numbers of radiological images at an exponentially growing rate, the medical field is far from a completely data-driven artificially intelligent diagnosis. The main reason is a lack of data harmony across multi-site studies, which keeps the training data substantially low for a truly large-scale study. Recognizing that homogenizing all CT acquisition parameters across practices may not be possible, our research goal is to provide a practical approach to this issue. Based on the data-mining approach we performed on our results from various parameters, our recommendations are as follows: 1) During volumetric measurements changes in slice thickness may produce acceptable errors, however the effects on texture are most dramatic. Hence, if possible, slice thickness should not be altered between studies if radiomic features are being compared between studies. 2) If scanning/reconstruction changes are inevitable, they should be limited to a single parameter. For example, only dose level or only sharpness should be changed (Fig. 8-10), and those changes should be kept to a minimum. Multiple parameter changes in general produces greater measurement errors (Fig. 4).

We performed a large-scale data mining approach for finding a "compatible" set of parameters, however, as it can be seen from the results compatibilities are very limited. In an earlier study, Young et al. [20] raised another concern pertaining to radiation dosage and kernel usage in CT lung-nodule volumetry; they found that the volume of lung nodules was extremely robust to the dose and reconstruction kernels. On the other hand, Chen et al. [8], [29]–[31] showed that the choice of reconstruction algorithm slightly affects the measurement of lesion volume. Regardless, reduced dosage comes at the cost of decreased image quality, which in turn make the results less reproducible for textures. Lo et al. [7] discussed these specific issues pertaining to lung CT and lung nodule quantification. Many authors recognize that it will be desirable to reduce the dose levels among the patients, and many vendors are already working towards quantitative imaging based on reduced dose levels [32]–[35]. In addition, other reports indicated that iterative reconstruction algorithms offer an opportunity to substantially reduce radiation dosage in case of CT scans while maintaining a good image resolution for visualization and nodule detections [8], [36]–[39]. However, the quantitative measurements from iterative algorithms can be significantly different from the traditional filtered back projection algorithms due to varying noise and resolution properties [40].

| Kernels | I26f | I31f | B31f | I40f | B40f | I50f | B50f | I70f | B60f | B70f |
|---|---|---|---|---|---|---|---|---|---|---|
| I26f | 100.00% | 54.70% | 36.26% | 70.55% | 47.12% | 19.51% | 12.88% | 8.32% | 4.84% | 4.22% |
| I31f | 54.70% | 100.00% | 49.93% | 52.87% | 37.64% | 16.65% | 9.03% | 6.25% | 4.23% | 3.73% |
| B31f | 36.26% | 49.93% | 100.00% | 37.39% | 51.76% | 21.36% | 10.87% | 7.42% | 4.29% | 3.92% |
| I40f | 70.55% | 52.87% | 37.39% | 100.00% | 53.90% | 21.21% | 13.18% | 8.39% | 4.73% | 4.13% |
| B40f | 47.12% | 37.64% | 51.76% | 53.90% | 100.00% | 26.45% | 14.62% | 9.54% | 5.76% | 4.90% |
| I50f | 19.51% | 16.65% | 21.36% | 21.21% | 26.45% | 100.00% | 22.09% | 12.17% | 7.26% | 5.65% |
| B50f | 12.88% | 9.03% | 10.87% | 13.18% | 14.62% | 22.09% | 100.00% | 23.92% | 10.35% | 8.21% |
| I70f | 8.32% | 6.25% | 7.42% | 8.39% | 9.54% | 12.17% | 23.92% | 100.00% | 18.60% | 14.90% |
| B60f | 4.84% | 4.23% | 4.29% | 4.73% | 5.76% | 7.26% | 10.35% | 18.60% | 100.00% | 37.66% |
| B70f | 4.22% | 3.73% | 3.92% | 4.13% | 4.90% | 5.65% | 8.21% | 14.90% | 37.66% | 100.00% |

Fig. 8. Percentage of compatible texture features in different kernel pairs while keeping slice thickness and dose levels fixed for all 23 patients and 28 features. For example, by changing the kernel from I31f to I26f, on average, 54.7% of the 28 texture features will be statistically the same (compatible) in our patient population under the same slice thickness and dose level configurations (32 different conditions).

| Thickness (mm) | 5.00 | 4.00 | 3.00 | 2.00 | 1.50 | 1.00 | 0.75 | 0.60 |
|---|---|---|---|---|---|---|---|---|
| 5.00 | 100.00% | 69.70% | 45.59% | 32.04% | 25.37% | 21.49% | 18.79% | 14.19% |
| 4.00 | 69.70% | 100.00% | 50.95% | 35.04% | 26.20% | 21.44% | 18.32% | 13.53% |
| 3.00 | 45.59% | 50.95% | 100.00% | 51.49% | 33.12% | 25.12% | 21.14% | 14.18% |
| 2.00 | 32.04% | 35.04% | 51.49% | 100.00% | 46.49% | 30.83% | 23.86% | 14.95% |
| 1.50 | 25.37% | 26.20% | 33.12% | 46.49% | 100.00% | 48.96% | 32.73% | 17.63% |
| 1.00 | 21.49% | 21.44% | 25.12% | 30.83% | 48.96% | 100.00% | 45.78% | 20.93% |
| 0.75 | 18.79% | 18.32% | 21.14% | 23.86% | 32.73% | 45.78% | 100.00% | 24.98% |
| 0.60 | 14.19% | 13.53% | 14.18% | 14.95% | 17.63% | 20.93% | 24.98% | 100.00% |

Fig. 9. Percentage of compatible texture features in different slice thickness pairs while keeping kernel and dose levels fixed for all 23 patients.

| Dose (%) | 100.00 | 50.00 | 25.00 | 12.50 |
|---|---|---|---|---|
| 100.00 | 100.00% | 52.61% | 34.54% | 22.30% |
| 50.00 | 52.61% | 100.00% | 43.47% | 25.96% |
| 25.00 | 34.54% | 43.47% | 100.00% | 34.82% |
| 12.50 | 22.30% | 25.96% | 34.82% | 100.00% |

Fig. 10. Percentage of compatible texture features in different dose level pairs while keeping kernel and slice thickness fixed for all 23 patients.

Based on earlier studies, high-resolution texture characterization requires image reconstruction using thin slices. However, thin slices also increase image noise; increasing slice thickness will decrease the noise while inducing blur. Based on our results, increasing slice thickness decreases the reproducibility of volumetric measurements but increases the reproducibility of histogram- and texture- based features. Thus, in this case, CT reconstruction becomes an optimization problem which corroborates our view based on these results that there might not be a best universal set of parameters that simultaneously covers both volumetric measures and radiomic features. Increasing slice thickness causes information loss due to the over-smoothing effect [41]. This may be the reason for increased compatibility for texture features.

Our methodology with full control of the reconstruction parameters had advantages and disadvantages. While limiting the number of scanners gave us the full advantage of image reconstruction algorithm compatibility and standardization, our results at this point are limited to few scanner types (Siemens Definition Flash, Definition AS, AS Plus, Edge). However, this approach also helped us to point out the potential issues in terms of CT image reconstruction. At this point, we believe that it would be even more difficult to show compatibilities among multiple scanners from multiple vendors.

Another limitation was our relatively small sample size (n = 23). This was due to fact that raw images can only be stored in our scanners for only 2 to 4 weeks (depending on the scanner and due to space restrictions). The study was initially designed as a retrospective analysis on nodule measurements, instead of classification of nodules. Under our circumstances, this would not be possible with retrospective analysis, and we would have needed to conduct a prospective study, which could have taken much longer time to complete.

Due to the small sample size, we only looked at nodules under 2~2.5 cm for their solid components, if any nodule had surrounding ground glass tissue, these components were ignored (two of our cases had minor ground glass features surrounding them), and only the solid components were included in the measurements and comparisons.

In conclusion, we found that slice thickness is the main factor impacting reproducibility of the image features we investigated. It is difficult to maintain both volumetric and radiomic measurement reproducibility and reliability simultaneously. However, our findings indicate that at a thickness of approximately 2mm volumetric measurement reproducibility can be maintained. However, especially for reproducibility in radiomic features, image scanning and reconstruction protocols need to remain stable. Standardization of the imaging acquisition parameters would become even more important in larger scale studies, where images are collected from multiple sites. As we have shown here, even with scanners with compatible image reconstruction parameters in a highly controlled environment, it is hard to maintain measurement reproducibility when parameters are arbitrarily changed. With images coming from multiple sites and multiple vendors, if studies are not designed and scanning protocols are not aligned properly; it can be very hard to produce reliable results that can be utilized within clinical studies.

TABLE I
EXTRACTED FEATURES

| Group | Feature | Definition |
|---|---|---|
| I. Histogram | 1. Mean | $\overline{X} = \frac{1}{N}\sum_{i=1}^{N} X(i)$ |
| | 2. Contrast | $\sqrt{\frac{1}{N}\sum_{i=1}^{N}(X(i) - \overline{X})^2}$ |
| | 3. Standard deviation | $\sqrt{\frac{1}{N-1}\sum_{i=1}^{N}(X(i) - \overline{X})^2}$ |
| | 4. Skewness | $\frac{\frac{1}{N}\sum_{i=1}^{N}(X(i)-\overline{X})^3}{\left(\sqrt{\frac{1}{N-1}\sum_{i=1}^{N}(X(i)-\overline{X})^2}\right)^3}$ |
| | 5. Kurtosis | $\frac{\frac{1}{N}\sum_{i=1}^{N}(X(i)-\overline{X})^4}{\left(\sqrt{\frac{1}{N}\sum_{i=1}^{N}(X(i)-\overline{X})^2}\right)^2}$ |
| II. GLCM | 6. Homogeneity (asm) | $\sum_i \sum_j \{p(i,j)\}^2$ |
| | 7. Contrast | $\sum_{n=0}^{N_g-1} n^2 \left\{\sum_{i=1}^{N_g} \sum_{j=1}^{N_g} p(i,j)\right\}$ |
| | 8. Correlation | $\frac{\sum_i \sum_j (ij)p(i,j) - \mu_x \mu_y}{\sigma_x \sigma_y}$ |
| | 9. Variance | $\sum_i \sum_j (i-\mu)^2 p(i,j)$ |
| | 10. Inverse difference moment | $\sum_i \sum_j \frac{1}{1+(1-j)^2} p(i,j)$ |
| | 11. Sum average | $\sum_{i=2}^{2N_g} i p_{x+y}(i)$ |
| | 12. Sum entropy | $-\sum_{i=2}^{2N_g} p_{x+y}(i) \log\{p_{x+y}(i)\}$ |
| | 13. Sum variance | $\sum_{i=2}^{2N_g} (i - \mu_{x+y})^2 p_{x+y}(i)$ |
| | 14. Entropy | $-\sum_i \sum_j p(i,j) \log(p(i,j))$ |
| | 15. Difference variance | $\sum_{i=0}^{N_g-1} (i - \mu_{x-y})^2 p_{x-y}(i)$ |
| | 16. Difference entropy | $-\sum_{i=0}^{N_g-1} p_{x-y}(i) \log\{p_{x-y}(i)\}$ |
| | 17. Measures of correlation1 | $\frac{Ent - HXY1}{max\{HX, HY\}}$ |
| | 18. Measures of correlation2 | $\sqrt[2]{1 - e^{-2(HXY2 - Ent)}}$ |
| III. RLM | 19. Short run emphasis | $\frac{\sum_{i=1}^{N_g} \sum_{j=1}^{N_r} \left[\frac{p(i,j|\theta)}{j^2}\right]}{\sum_{i=1}^{N_g} \sum_{j=1}^{N_r} p(i,j|\theta)}$ |
| | 20. Long run emphasis | $\frac{\sum_{i=1}^{N_g} \sum_{j=1}^{N_r} j^2 p(i,j|\theta)}{\sum_{i=1}^{N_g} \sum_{j=1}^{N_r} p(i,j|\theta)}$ |
| | 21. Grey level non-uniformity | $\frac{\sum_{i=1}^{N_g} \left[\sum_{j=1}^{N_r} p(i,j|\theta)\right]^2}{\sum_{i=1}^{N_g} \sum_{j=1}^{N_r} p(i,j|\theta)}$ |
| | 22. Run length non-uniformity | $\frac{\sum_{j=1}^{N_r} \left[\sum_{i=1}^{N_g} p(i,j|\theta)\right]^2}{\sum_{i=1}^{N_g} \sum_{j=1}^{N_r} p(i,j|\theta)}$ |
| | 23. Run percentage | $\sum_{i=1}^{N_g} \sum_{j=1}^{N_r} \frac{p(i,j|\theta)}{N_p}$ |
| IV. NGLDM | 24. Small Number Emphasis | $\sum_{k=1}^{K} \sum_{s=1}^{S} [Q(k,s)/s^2] / \sum_{k=1}^{K} \sum_{s=1}^{S} Q(k,s)$ |
| | 25. Large Number Emphasis | $\sum_{k=1}^{K} \sum_{s=1}^{S} [s^2 Q(k,s)] / \sum_{k=1}^{K} \sum_{s=1}^{S} Q(k,s)$ |
| V. NGTDM | 26. Coarseness | $\left[\sum_{i=0}^{L_h} p_i s(i)\right]^{-1}$ |
| | 27. Complexity | $\sum_{i=0}^{L_h} \sum_{j=0}^{L_h} \{(\|i-j\|)/(n^2(p_i + p_j))\}\{p_i s(i) + p_j s(j)\}$ |
| | 28. Texture Strength | $\frac{\left[\sum_{i=0}^{L_h} \sum_{j=0}^{L_h} (p_i + p_j)(i-j)^2\right]}{\left[\sum_{i=0}^{L_h} s(i)\right]}$ |

## References

[1] A. A. Bankier, H. MacMahon, J. M. Goo, G. D. Rubin, C. M. Schaefer-Prokop, and D. P. Naidich, "Recommendations for measuring pulmonary nodules at CT: A statement from the Fleischner Society," *Radiology*, vol. 285, no. 2, pp. 584–600, 2017.

[2] M. A. Heuvelmans, J. E. Walter, R. Vliegenthart, P. M. A. van Ooijen, G. H. De Bock, H. J. De Koning et al., "Disagreement of diameter and volume measurements for pulmonary nodule size estimation in CT lung cancer screening," *Thorax*, vol. 73, no. 8, pp. 779–781, 2018.

[3] H. J. W. L. Aerts, E. R. Velazquez, R. T. H. Leijenaar, C. Parmar, P. Grossmann, S. Carvalho et al., "Decoding tumour phenotype by noninvasive imaging using a quantitative radiomics approach," *Nat. Commun.*, vol. 5, p. 4006, 2014.

[4] R. J. Gillies, P. E. Kinahan, and H. Hricak, "Radiomics: Images are more than pictures, they are data," *Radiology*, vol. 278, no. 2, pp. 563–577, 2015.

[5] B. Zhao, Y. Tan, W.-Y. Tsai, J. Qi, C. Xie, L. Lu et al., "Reproducibility of radiomics for deciphering tumor phenotype with imaging," *Sci. Rep.*, vol. 6, p. 23428, 2016.

[6] K. Buch, B. Li, M. M. Qureshi, H. Kuno, S. W. Anderson, and O. Sakai, "Quantitative assessment of variation in CT parameters on texture features: Pilot study using a nonanatomic phantom," *Am. J. Neuroradiol.*, vol. 38, no. 5, pp. 981–985, 2017.

[7] P. Lo, S. Young, H. J. Kim, M. S. Brown, and M. F. McNitt-Gray, "Variability in CT lung-nodule quantification: Effects of dose reduction and reconstruction methods on density and texture based features," *Med. Phys.*, vol. 43, no. 8Part1, pp. 4854–4865, 2016.

[8] B. Chen, J. C. Ramirez Giraldo, J. Solomon, and E. Samei, "Evaluating iterative reconstruction performance in computed tomography," *Med. Phys.*, vol. 41, no. 12, p. 121913, 2014.

[9] K. Buch, A. Fujita, B. Li, Y. Kawashima, M. M. Qureshi, and O. Sakai, "Using texture analysis to determine human papillomavirus status of oropharyngeal squamous cell carcinomas on CT," *Am. J. Neuroradiol.*, vol. 36, no. 7, pp. 1343–1348, 2015.

[10] X. Fave, M. Cook, A. Frederick, L. Zhang, J. Yang, D. Fried et al., "Preliminary investigation into sources of uncertainty in quantitative imaging features," *Comput. Med. Imaging Graph.*, vol. 44, pp. 54–61, 2015.

[11] D. V Fried, S. L. Tucker, S. Zhou, Z. Liao, O. Mawlawi, G. Ibbott et al., "Prognostic value and reproducibility of pretreatment CT texture features in stage III non-small cell lung cancer," *Int. J. Radiat. Oncol. Biol. Phys.*, vol. 90, no. 4, pp. 834–842, 2014.

[12] B. Ganeshan and K. A. Miles, "Quantifying tumour heterogeneity with CT," *Cancer Imaging*, vol. 13, no. 1, pp. 140–149, 2013.

[13] K. A. Miles, B. Ganeshan, M. R. Griffiths, R. C. D. Young, and C. R. Chatwin, "Colorectal cancer: Texture analysis of portal phase hepatic CT images as a potential marker of survival," *Radiology*, vol. 250, no. 2, pp. 444–452, 2009.

[14] G. D. Rubin, "Lung nodule and cancer detection in computed tomography screening," *J. Thorac. Imaging*, vol. 30, no. 2, pp. 130–138, 2015.

[15] F. Fischbach, F. Knollmann, V. Griesshaber, T. Freund, E. Akkol, and R. Felix, "Detection of pulmonary nodules by multislice computed tomography: Improved detection rate with reduced slice thickness," *Eur. Radiol.*, vol. 13, no. 10, pp. 2378–2383, 2003.

[16] M. Sinsuat, S. Saita, Y. Kawata, N. Niki, H. Ohmatsu, T. Tsuchida et al., "Influence of slice thickness on diagnoses of pulmonary nodules using low-dose CT: Potential dependence of detection and diagnostic agreement on features and location of nodule," *Acad. Radiol.*, vol. 18, no. 5, pp. 594–604, 2011.

[17] E. A. Kazerooni, J. H. Austin, W. C. Black, D. S. Dyer, T. R. Hazelton, A. N. Leung et al., "ACR-STR practice parameter for the performance and reporting of lung cancer screening thoracic computed

tomography (CT): 2014 (Resolution 4)," *J. Thorac. Imaging*, vol. 29, no. 5, pp. 310–316, 2014.

[18] C. A. Gatsonis, D. R. Aberle, C. D. Berg, W. C. Black, T. R. Church, R. M. Fagerstrom *et al.*, "The national lung screening trial: Overview and study design," *Radiology*, vol. 258, no. 1, pp. 243–253, 2011.

[19] A. P. Reeves, A. M. Biancardi, T. V Apanasovich, C. R. Meyer, H. MacMahon, E. J. R. van Beek *et al.*, "The Lung Image Database Consortium (LIDC): A comparison of different size metrics for pulmonary nodule measurements," *Acad. Radiol.*, vol. 14, no. 12, pp. 1475–1485, 2007.

[20] S. Young, H. J. G. Kim, M. M. Ko, W. W. Ko, C. Flores, and M. F. McNitt-Gray, "Variability in CT lung-nodule volumetry: Effects of dose reduction and reconstruction methods," *Med. Phys.*, vol. 42, no. 5, pp. 2679–2689, 2015.

[21] P. Therasse, S. G. Arbuck, E. A. Eisenhauer, J. Wanders, R. S. Kaplan, L. Rubinstein *et al.*, "New guidelines to evaluate the response to treatment in solid tumors," *J. Natl. Cancer Inst.*, vol. 92, no. 3, pp. 205–216, 2000.

[22] N. Petrick, H. J. G. Kim, D. Clunie, K. Borradaile, R. Ford, R. Zeng *et al.*, "Comparison of 1D, 2D, and 3D nodule sizing methods by radiologists for spherical and complex nodules on thoracic CT phantom images," *Acad. Radiol.*, vol. 21, no. 1, pp. 30–40, 2014.

[23] J.-M. Kuhnigk, V. Dicken, L. Bornemann, D. Wormanns, S. Krass, and H.-O. Peitgen, "Fast automated segmentation and reproducible volumetry of pulmonary metastases in CT-scans for therapy monitoring," in *International Conference on Medical Image Computing and Computer-Assisted Intervention*, 2004, pp. 933–941.

[24] L. R. Goodman, M. Gulsun, L. Washington, P. G. Nagy, and K. L. Piacsek, "Inherent variability of CT lung nodule measurements in vivo using semiautomated volumetric measurements," *Am. J. Roentgenol.*, vol. 186, no. 4, pp. 989–994, 2006.

[25] M.-P. Revel, A. Merlin, S. Peyrard, R. Triki, S. Couchon, G. Chatellier *et al.*, "Software volumetric evaluation of doubling times for differentiating benign versus malignant pulmonary nodules," *Am. J. Roentgenol.*, vol. 187, no. 1, pp. 135–142, 2006.

[26] A. P. Reeves, A. B. Chan, D. F. Yankelevitz, C. I. Henschke, B. Kressler, and W. J. Kostis, "On measuring the change in size of pulmonary nodules," *IEEE Trans. Med. Imaging*, vol. 25, no. 4, pp. 435–450, 2006.

[27] B. Zhao, Y. Tan, D. J. Bell, S. E. Marley, P. Guo, H. Mann *et al.*, "Exploring intra-and inter-reader variability in uni-dimensional, bi-dimensional, and volumetric measurements of solid tumors on CT scans reconstructed at different slice intervals," *Eur. J. Radiol.*, vol. 82, no. 6, pp. 959–968, 2013.

[28] M. A. Gavrielides, L. M. Kinnard, K. J. Myers, and N. Petrick, "Noncalcified lung nodules: Volumetric assessment with thoracic CT," *Radiology*, vol. 251, no. 1, pp. 26–37, 2009.

[29] B. Chen, D. Marin, S. Richard, D. Husarik, R. Nelson, and E. Samei, "Precision of iodine quantification in hepatic CT: Effects of iterative reconstruction with various imaging parameters," *Am. J. Roentgenol.*, vol. 200, no. 5, pp. W475–W482, 2013.

[30] B. Chen, O. Christianson, J. M. Wilson, and E. Samei, "Assessment of volumetric noise and resolution performance for linear and nonlinear CT reconstruction methods," *Med. Phys.*, vol. 41, no. 7, p. 71909, 2014.

[31] B. Chen, H. Barnhart, S. Richard, M. Robins, J. Colsher, and E. Samei, "Volumetric quantification of lung nodules in CT with iterative reconstruction (ASiR and MBIR)," *Med. Phys.*, vol. 40, no. 11, p. 111902, 2013.

[32] L. A. Hunter, S. Krafft, F. Stingo, H. Choi, M. K. Martel, S. F. Kry *et al.*, "High quality machine-robust image features: Identification in nonsmall cell lung cancer computed tomography images," *Med. Phys.*,

vol. 40, no. 12, p. 121916, 2013.

[33] D. Mackin, X. Fave, L. Zhang, D. Fried, J. Yang, B. Taylor *et al.*, "Measuring computed tomography scanner variability of radiomics features," *Invest. Radiol.*, vol. 50, no. 11, pp. 757–765, 2015.

[34] X. Fave, D. Mackin, J. Yang, J. Zhang, D. Fried, P. Balter *et al.*, "Can radiomics features be reproducibly measured from CBCT images for patients with non-small cell lung cancer?," *Med. Phys.*, vol. 42, no. 12, pp. 6784–6797, 2015.

[35] M. J. Nyflot, F. Yang, D. Byrd, S. R. Bowen, G. A. Sandison, and P. E. Kinahan, "Quantitative radiomics: Impact of stochastic effects on textural feature analysis implies the need for standards," *J. Med. Imaging*, vol. 2, no. 4, p. 41002, 2015.

[36] M. Beister, D. Kolditz, and W. A. Kalender, "Iterative reconstruction methods in X-ray CT," *Phys. Medica*, vol. 28, no. 2, pp. 94–108, 2012.

[37] J.-B. Thibault, K. D. Sauer, C. A. Bouman, and J. Hsieh, "A three-dimensional statistical approach to improved image quality for multislice helical CT," *Med. Phys.*, vol. 34, no. 11, pp. 4526–4544, 2007.

[38] J. Solomon, A. Mileto, J. C. Ramirez-Giraldo, and E. Samei, "Diagnostic performance of an advanced modeled iterative reconstruction algorithm for low-contrast detectability with a third-generation dual-source multidetector CT scanner: Potential for radiation dose reduction in a multireader study," *Radiology*, vol. 275, no. 3, pp. 735–745, 2015.

[39] P. J. Pickhardt, M. G. Lubner, D. H. Kim, J. Tang, J. A. Ruma, A. M. del Rio *et al.*, "Abdominal CT with model-based iterative reconstruction (MBIR): Initial results of a prospective trial comparing ultralow-dose with standard-dose imaging," *Am. J. Roentgenol.*, vol. 199, no. 6, pp. 1266–1274, 2012.

[40] B. Schulz, M. Beeres, B. Bodelle, R. Bauer, F. Al-Butmeh, A. Thalhammer *et al.*, "Performance of iterative image reconstruction in CT of the paranasal sinuses: A phantom study," *Am. J. Neuroradiol.*, vol. 34, no. 5, pp. 1072–1076, 2013.

[41] B. Zhao, Y. Tan, W. Y. Tsai, L. H. Schwartz, and L. Lu, "Exploring variability in CT characterization of tumors: A preliminary phantom study," *Transl. Oncol.*, vol. 7, no. 1, pp. 88–93, 2014.